\documentclass{article}

\usepackage{hyperref}       % hyperlinks
\usepackage{url}            % simple URL typesetting
\usepackage{booktabs}       % professional-quality tables
\usepackage{amsfonts}       % blackboard math symbols
\usepackage{nicefrac}       % compact symbols for 1/2, etc.
\usepackage{microtype}      % microtypography
\usepackage{graphicx}

\newtheorem{conjecture}{Conjecture}[section]

\begin{document}
\title{Designing communication systems via iterative improvement: error correction coding with Bayes decoder and codebook optimized for source symbol error}

\author{
  Chai Wah Wu\\
  IBM Research AI\\
  IBM T. J.~Watson Research Center\\
  P. O. Box 218\\
  Yorktown Heights, NY 10598 \\
  \texttt{cwwu@us.ibm.com} \\
}

\date{April 23, 2019\\ Latest update: October 7, 2021}

\maketitle

\begin{abstract}
In most error correction coding (ECC) frameworks, the typical error metric is the bit error rate (BER) which measures the number of bit errors. 
For this metric, the positions of the bits are not relevant to the decoding, and in many noise models, not relevant to the BER either. In many applications
this is unsatisfactory as typically all bits are not equal and have different significance.  We consider the problem of bit error correction and mitigation where bits in different positions have different importance. For error correction, we look at ECC from a Bayesian perspective 
and introduce Bayes estimators with general loss functions to take into account the bit significance. We propose
ECC schemes that optimize this error metric. As the problem is highly nonlinear, traditional ECC construction techniques are not applicable. Using exhaustive search is cost prohibitive, and thus we use iterative improvement search techniques to find good codebooks. We optimize both general codebooks and linear codes. We provide numerical experiments to show that they can be superior to classical linear block codes such as Hamming codes and decoding methods such as minimum distance decoding.

For error mitigation, we study the case where ECC is not possible or not desirable, but significance aware encoding of information is still beneficial in reducing the average error. We propose a novel number presentation format suitable for emerging storage media where the noise magnitude is unknown and possibly large and show that it has lower mean error than the traditional number format.
\end{abstract}

\section{Introduction} \label{sec:intro}

The information bit error rate (BER) in classical error coding  is based on the Hamming distance, i.e. the number of bits that are different between the symbols at the transmitter and the decoded symbols at the receiver.
For this metric, the positions of the source bits where the error occurred are not significant. 
This is unsatisfactory since in many scenarios all source bits are not equal and have different significance. For instance, representing an integer in binary, the bits will have different significance with the difference increasing exponentially with its position, and an error in the most significant bit is much more problematic than an error in the least significant bit. Furthermore, the relationship between the difference $|i-j|$ of two integers $i$ and $j$ and the Hamming distance when expressed as bitstrings is nonlinear and nonmonotonic. For instance, the numerical difference between the numbers $8$ and $7$ is $1$, but expressed as bits, their Hamming distance is $4$.  On the other hand, the difference between $0$ and $2^k$ is $2^k$, but have a Hamming distance of $1$. In image compression \cite{pennebaker:jpeg1993}, the discrete cosine transform (DCT) coefficients for lower frequencies are more important than the higher frequencies coefficients and luminance coefficients are more important than chrominance coefficients as the human visual system exhibits low-pass behavior and is more sensitive to luminance changes.  In stored-program computers, where both program data and instruction data are stored in memory, the program instruction code is more critical than program data as an erroneous instruction can cause the machine to crash whereas incorrect data typically leads to incorrect results, but not cause a machine to crash \cite{Stefanakis2014}.

The purpose of this paper is to consider some approaches to take into account the difference in significance of the source bits in the context of communication systems or storage systems where error correcting codes and source coding are used to combat channel and storage noise.

\section{Notation and setup}\label{sec:setup}
\begin{figure}[htbp]
\centerline{\includegraphics[width=4.3in]{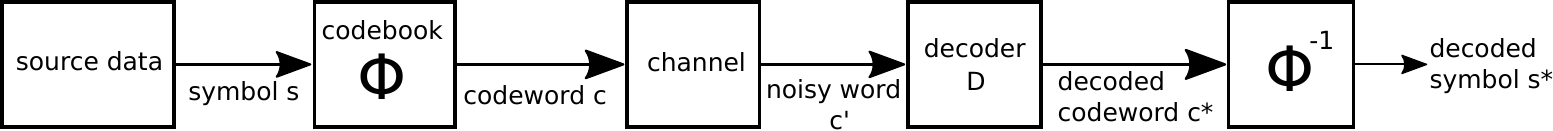}}
\caption{Communication system setup}\label{fig:framework}
\end{figure}

For an integer in the range $0 \leq n < 2^k$, let $b_k(n)$ denote the $k$-bit representation of $n$, e.g. $b_4(9) = 1001$.
Let us denote the bijection between symbols $s\in S$ (also called {\em information} or {\em source} symbols) and codewords $c\in C$ by $\Phi$. 
We assume that each symbol in $S$ occurs with equal probability in the data stream, i.e. $\forall s \in S, p(s) = \frac{1}{|S|}$. The standard setup \cite{haykin:comm:2009} is shown in Fig. \ref{fig:framework}, where each symbol $s$ is mapped to the corresponding codeword $c = \Phi(s)$  via the codebook $\Phi$ and transmitted through the channel\footnote{As we focus on ECC, we will ignore channel modulation/demodulation for now.}. The act of the codewords moving through the channel incurs errors because of the noisy channel. This channel can be physical space in the case of communication channels or time in the case of storage systems. At the receiver, the received noisy word $c'$ is decoded as $D(c') = c^* \in C$ and the decoded symbol is retrieved\footnote{The mapping $\Phi^{-1}$ can be efficiently implemented via a hash table which is indexed by codewords and maps codewords to symbols.} as $s^* = \Phi^{-1}(c^*)$.

\section{Ideal observer, maximum likelihood and minimum distance decoding}

Given a transmitted codeword $c$ and a received word $c'$ (that may or may not be a codeword in $C$), the ideal observer decoding \cite{roman:code:1997} returns the codeword $c^*$ such that
$c^* = \mbox{argmax}_{w\in C} p(w|c')$ where $p(w|c')$ is the probability that codeword $w$ is sent given that $c'$ (which is not necessarily a codeword) is received. Given our assumption of a uniform prior distribution for codewords $c$, this is equivalent to maximum likelihood decoding.  For certain channel models (e.g. binary symmetric channel with $p < 0.5$), this is equivalent to minimum distance decoding.

\section{Error rate in source symbol space}
To take into account the difference in significance of the bits of the source symbols, instead of BER, we need an error metric that measures the difference between source symbols. Let us consider the following error rate:
$e_\delta = \lim_{N\rightarrow\infty} \frac{1}{N}\sum_{i=1}^N \delta(s(i),s^*(i))$ where $s(i)$ is the $i$-th source symbol and $s^*(i)$ is the $i$-th decoded symbol. The function $\delta$ measures the difference between the transmitted symbol and the decoded symbol in the source symbol space. 
In our example of comparing integers, we define the following two functions: $\delta_1 = |b_k^{-1}(s)-b_k^{-1}(s^*)|$ and  $\delta_2 = (b_k^{-1}(s)-b_k^{-1}(s^*))^2$  , i.e. the difference (resp. squared difference) between $s$ and $s^*$ when expressed as integers.

In order to minimize this error rate, we optimize both the decoding method $D$ and the codebook $\Phi$. We will look at the optimization of each of these in turn.

\section{Bayes estimator}  \label{sec:bayes}
In Bayesian estimation theory \cite{keener:stats:2010}, the Bayes estimator
$c^*$ minimizes the expected loss $E(L(s^*,s)|c')$, where $L(s^*,s)$ is a {\em loss function} measuring the difference between the decoded source symbol $s^*$ and the transmitted source symbol $s$. For example, let $a_i(x)$ denote the $i$-th bit of $x$, then one possible loss function is $L(s^*,s) = \sum_i \alpha_i |a_i(s^*)-a_i(s)|$. If we choose $\alpha_i = 2^i$, then $L(s^*,s)$ is equal to the numerical value of the bitwise XOR of $s^*$ and $s$.

Consider the posterior distribution $p(s|c')$. For the additive white Gaussian noise (AWGN) channel, $p(s|c')$ is proportional to $g_{(\mu,\sigma)}(c'-\Phi(s))$ where $g$ is the multivariate Gaussian pdf with mean $\mu$ and variance $\sigma^2$.  For a general loss function on a finite discrete symbol space, the Bayes estimator can be implemented by comparing all possible codebook candidates given the received word. For the case when $s$ is a scalar quantity and for some special forms of the loss function,  
the following simple explicit forms of the Bayes estimator are well known and lead to efficient implementations of the Bayes estimator:
\begin{itemize}
\item If the loss function is the 0-1 loss function, i.e. $L(s^*,s) = 0$ if $s^* = s$ and $1$ otherwise, then the Bayes estimator is the mode of the posterior distribution and is equivalent to the maximum-a-posteriori estimate or the ideal observer.
\item If the loss function is the squared error loss function, i.e. $L(s^*,s) = (s^*-s)^2$, then the Bayes estimator is the mean of the posterior distribution.
\item If the loss function is the absolute value loss function, i.e. $L(s^*,s) = |s^*-s|$, then the Bayes estimator is the median of the posterior distribution.
\end{itemize}

The Bayes risk is defined as $E_\pi (L(s^*,s))$, where $\pi$ is the prior distribution of $s$ and the Bayes estimator is the estimator that minimizes the Bayes risk. Since $\pi$ is the uniform distribution by our assumption, $\frac{1}{N}\sum_i L(s^*(i),s(i))$ is an unbiased estimator of the Bayes risk. Thus the Bayes estimator is the appropriate choice in order to minimize $\frac{1}{N}\sum_i L(s^*(i),s(i))$, which is equal to $e_\delta$ if
\begin{equation}\label{eqn:loss}
L(s^*,s) = \delta(s,s^*)\end{equation} 
This analysis shows that we should choose the loss function according to Eq. (\ref{eqn:loss}).

\section{An iterative improvement approach to finding good codebooks}
In the previous section we have determined the optimal decoding scheme for a specific $e_\delta$ and a specific codebook. Next we need to find an optimal codebook $\Phi$ that minimizes $e_\delta$ further. While Shannon's channel coding theorems show the existence of an optimal codebook asymptotically, finding a good codebook for arbitrary lengths has proven to be a difficult task.  If the error metric is for instance a Hamming distance in the codeword space, then traditional coding constructions (e.g. Hamming codes, block codes) can generate codebooks efficiently via linear algebraic techniques, although ther performance is poor as the codeword length grows. In fact, even for the Hamming distance metric, a general construction of optimal code does not exist.  In our case, the error is a general error in the source symbol space and such techniques are not applicable anymore.  On the other hand, solving it via exhaustive search is not feasible for codebooks of large lengths.

 For many optimization problems where there is no gradient or the gradient is difficult to compute,  gradient-based nonlinear programming algorithms are not applicable. In this case, AI-based optimization heuristics have proven to be useful to find good and near-optimal solutions to such problems in much less time than an exhaustive search would entail. In \cite{haas:ecc:2007} such techniques have been used to find codebooks minimizing the Hamming distance for small codebooks.  One of the contributions of this paper is to explore whether such AI-based optimization methods can be used to construct a good codebook based on the bit-dependent error metric.

To this end, we use the following heuristic. We find a codebook that
minimizes the objective function $v = \sum_{i\neq j} \delta(s_i,s_j)p_c(\Phi(s_i)|\Phi(s_j))$ where $p_c(x|y)$ for $x, y \in C$ is the probability of receiving codeword $x$ when codeword $y$ is transmitted. The reason for choosing $v$ in this form is that for a fixed symbol $s$, the quantity $\sum_{s_i\neq s} \delta(s_i,s)p_c(\Phi(s_i)|\Phi(s))$ is an estimate of the mean error between the transmitted symbol and received symbol when $s$ is transmitted by sampling only on the codebook and thus $v$ is an estimate of the mean error when the source symbols are transmitted with equal probability.
For the AWGN channel with  zero mean and  variance $\sigma^2$, $p_c(x|y)$ is proportional to $e^{-\frac{d_H(x,y)}{2\sigma^2}}$ where $d_H$ is the Hamming distance. Thus we get  

$$ v =  \sum_{i\neq j} \delta(s_i,s_j)e^{-\frac{ d_H(\Phi(s_i),\Phi(s_j)))}{2\sigma^2}}.$$

We use several search algorithms including genetic algorithm \cite{genetic:goldberg:1989}, and various types of hill-climbing algorithms \cite{russell:ai:2010} to find codebooks that minimizes $v$. To ensure the codewords are distinct, we add a penalty term to the fitness or objective function whenever a candidate codebook satisfies $d_H(\Phi(s_i),\Phi(s_j)) = 0$ for some $i\neq j$. Note that the optimal notebook depends on the error function $e_\delta$ and the channel probability $p_c$.
For the genetic algorithm, the codebook as a whole can be represented as a binary string forming a chromosome. Our experiments found that a simple genetic algorithm with one point crossover and swap mutation performs better than hill-climbing in minimizing $v$.

\section{Comparison with related work}
In unequal error protection (UEP), different classes of messages are assigned different priorities and different ECC methods are applied to each class \cite{Borade2009,trott:uep:1996}. While there are similarities in the sense that data of higher importance should have more protection against communication errors, there are some differences between UEP and the current approach. 
In particular, in most prior UEP frameworks, the data is classified into multiple priority classes and different ECC schemes are applied independently to each class and the errors between the different classes do not trade off against each other.
On the other hand, in the current approach the bits have different significance and they all contribute to the same objective (e.g. contributing to the value of an integer) and thus their error correction should be treated holistically. To do this, we formulate an optimization problem where the objective function combines the various bits with different significances and thus such a trade off is possible. 
Furthermore, a Bayes optimal decoder is proposed here which for certain objective functions has a simple form and implementation.

While ML approaches have been applied to classical ECC code design (e.g.  \cite{haas:ecc:2007}), it has not been applied to code design with different bit significance. Also, in contrast to  \cite{haas:ecc:2007}, the current approach can be a data-driven approach and extendable to use information extracted from empirical channel data (see Section \ref{sec:results}).

\section{Numerical results} \label{sec:results}
In this section, we illustrate this framework by comparing the optimized codebooks and Bayes decoding with some well known linear block codes and decoding methods. To show that this error metric is different from the traditional Hamming distance metric, we show that codes optimized for this metric perform better than Hamming codes under this metric. Note that Hamming codes are perfects codes achieving the Hamming bound and thus are optimal under the Hamming distance metric.

\subsection{Block code of rate 4/7} \label{sec:example1}
In this case $k = 4$. Let the symbols be $0,1,\dots, 15$ and each symbol is mapped to a length 7 binary string.
We constructed an optimized codebook assuming an AWGN channel with $\sigma=1$. We simulate the performance of this codebook and that of the Hamming (7,4) code \cite{cover:information:1991}. We choose Hamming codes as a baseline as they are perfect codes and achieve the Hamming sphere-packing bound. However, as our experiments show, even though the Hamming code is optimal with respect to the Hamming distance, it is not optimal when considering other types of error metrics.  We consider 3 decoding schemes:
\begin{enumerate}
\item {\bf Hard decoding}: the received signal is quantized to a binary word and the decoded codeword is the closest codeword with respect to the Hamming distance.
\item {\bf Soft decoding}: the decoded codeword is the nearest codeword with respect to the Euclidean distance.
\item {\bf Bayes decoding}, as described above. For an AWGN channel the received noisy word is $c' = c + n$ and the Bayes estimator requires knowing (or estimating) the variance of the noise. With the assumption that the codewords are equiprobable, the variance of the codewords bits can be easily computed as the population variance of the bits of codebook $\Phi$. Since the noise is uncorrelated with the codewords, the variance of the noise can be estimated by subtracting the variance of the codewords bits from the estimated variance of the received noisy bits. 
\end{enumerate}

We consider both error metrics $e_{\delta_1}$ and $e_{\delta_2}$. As described earlier, the Bayes estimator is simply the median and mean of the posterior distribution in the symbol space $S$ respectively.
Furthermore, the codebook used is optimized by minimizing $v$ using a genetic algorithm for $e_{\delta_1}$ resp. $e_{\delta_2}$. In particular, for $e_{\delta_1}$, an optimized codebook found by the genetic algorithm is:

$$\left(\begin{array}{ccccccc}0 & 1 & 1 & 0 & 0 & 0 & 0\\0 & 1 & 0 & 0 & 0 & 0 & 0\\0 & 0 & 0 & 0 & 0 & 0 & 0\\0 & 0 & 0 & 0 & 0 & 0 & 1\\0 & 0 & 0 & 1 & 0 & 0 & 0\\0 & 1 & 0 & 1 & 0 & 1 & 0\\0 & 1 & 0 & 1 & 1 & 1 & 0\\1 & 1 & 0 & 1 & 1 & 0 & 0\\1 & 1 & 0 & 1 & 1 & 0 & 1\\1 & 1 & 1 & 1 & 1 & 0 & 1\\1 & 1 & 1 & 1 & 0 & 1 & 1\\1 & 0 & 1 & 1 & 0 & 1 & 1\\0 & 0 & 1 & 1 & 1 & 1 & 1\\1 & 0 & 1 & 1 & 1 & 1 & 1\\1 & 0 & 1 & 0 & 1 & 1 & 1\\1 & 0 & 1 & 0 & 1 & 1 & 0\end{array}\right)$$
where each row is a codeword and the first codeword corresponds to source symbol 0 and the last codeword corresponds to source symbol 15. Because of the additive white noise assumption, permuting the columns of this codebook will not affect the error rate. On the other hand, permuting the rows of the codebook (which corresponds to permuting the codewords) will affect the error rate. This is in contrast to the Hamming distance error metric (for which the Hamming code is optimized for) which does not change under permutation of the codewords.  

\begin{figure}[htbp]
\centerline{\includegraphics[width=5in]{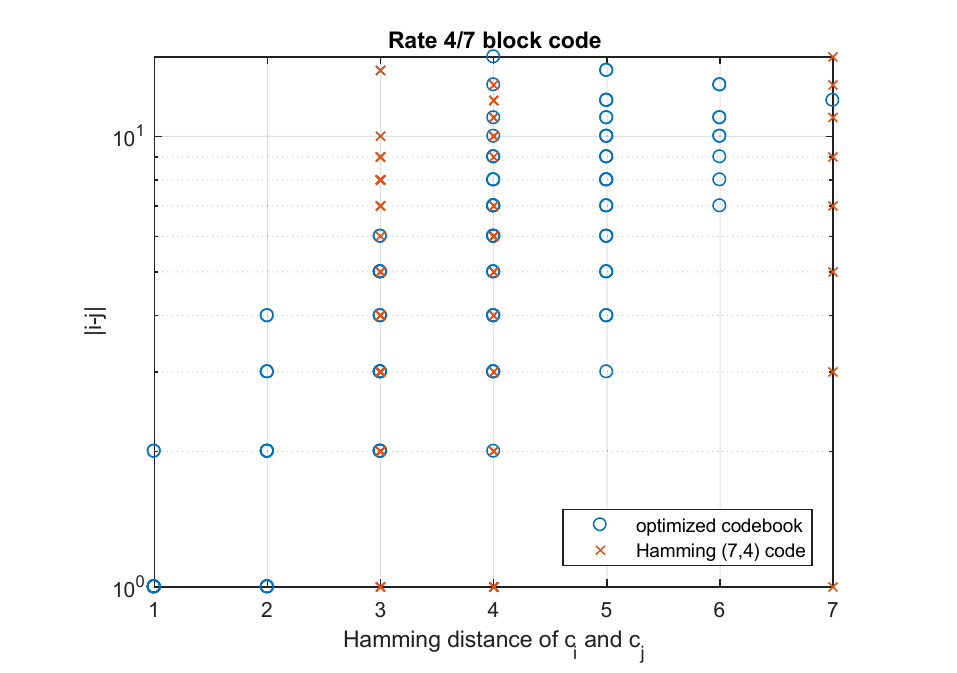}}
\caption{Hamming distance of codewords of $i$ and $j$ versus $|i-j|$ for rate 4/7.}\label{fig:l1-codebook}
\end{figure}

As the error metric measures the difference between symbols as integers, the Hamming distance of the codewords in this optimized codebook are correlated with the difference of the symbols as integers, not as bitstrings. For instance, recall that $7$ and $8$ differ by $1$ as integers but has a Hamming distance of $4$ as bitstrings. In this codebook the Hamming distance between the codewords for $7$ and $8$ is $1$. Similarly, $8$ and $0$ differ by $8$ as integers but have a Hamming distance of $1$ as bitstrings. In the codebook, the codewords for $0$ and $8$ have a Hamming distance of $5$. This is further illustrated in Fig. \ref{fig:l1-codebook}, where we show a plot of the Hamming distance of the codewords $c_i$ and $c_j$ for $i$ and $j$ versus $|i-j|$.  Ideally, we want a linear relationship between the Hamming distance of the codewords for $i$ and $j$ and $\log |i-j|$, i.e., numbers that are further apart numerically should have corresponding codewords with larger Hamming distance. We see that the optimized codebook does a much better job at satisfying this requirement than the Hamming 
$(7,4)$ code whose codewords only have a pairwise Hamming distance of  $3$, $4$ or $7$.

For $e_{\delta_2}$, an optimized codebook is:

$$\left(\begin{array}{ccccccc} 0 & 0 & 1 & 0 & 1 & 1 & 1\\0 & 0 & 1 & 0 & 0 & 1 & 1\\1 & 0 & 1 & 0 & 1 & 1 & 1\\0 & 0 & 1 & 0 & 1 & 1 & 0\\1 & 0 & 1 & 0 & 0 & 1 & 1\\1 & 0 & 1 & 0 & 0 & 1 & 0\\1 & 0 & 1 & 1 & 0 & 1 & 0\\1 & 0 & 0 & 0 & 0 & 1 & 0\\1 & 1 & 1 & 0 & 0 & 0 & 0\\1 & 1 & 0 & 0 & 0 & 0 & 1\\1 & 1 & 0 & 1 & 0 & 0 & 1\\1 & 1 & 0 & 1 & 0 & 0 & 0\\1 & 1 & 0 & 1 & 1 & 0 & 1\\1 & 1 & 0 & 1 & 1 & 0 & 0\\0 & 1 & 0 & 1 & 0 & 0 & 0\\0 & 1 & 0 & 1 & 1 & 0 & 0\end{array}\right)$$

\begin{figure}[htbp]
\centerline{\includegraphics[width=5in]{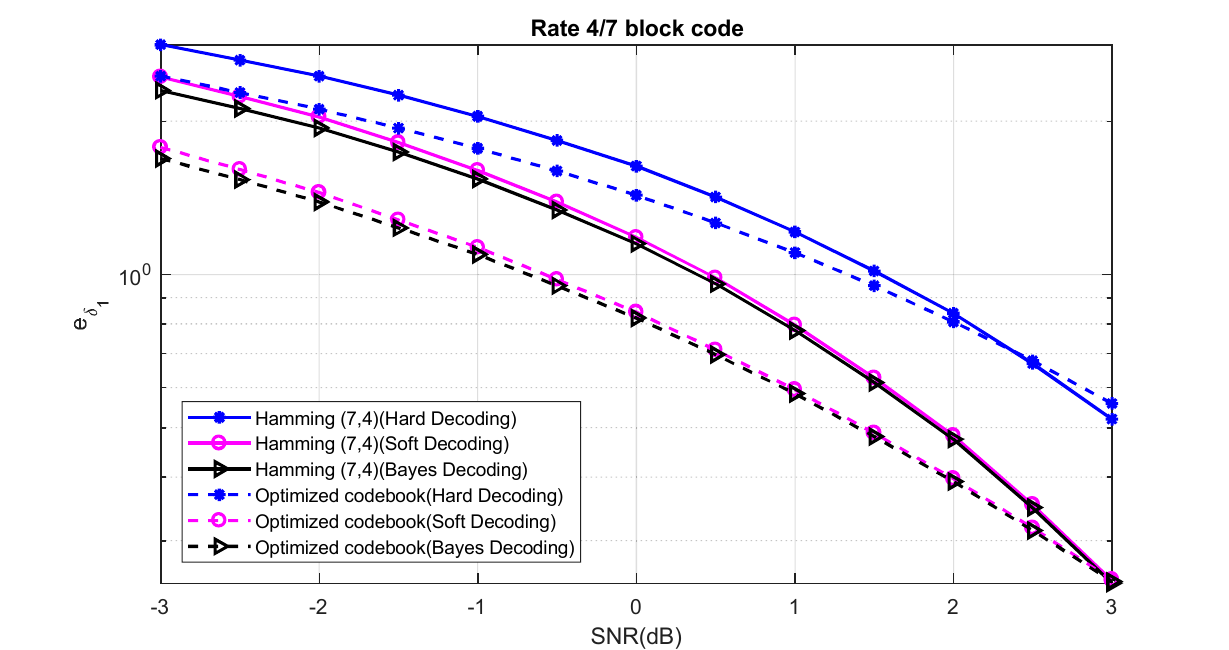}}
\caption{$e_{\delta_1}$ versus SNR for the rate 4/7 optimized code compared with Hamming (7,4) code.}\label{fig:hamming-l1}
\end{figure}

\begin{figure}[htbp]
\centerline{\includegraphics[width=5in]{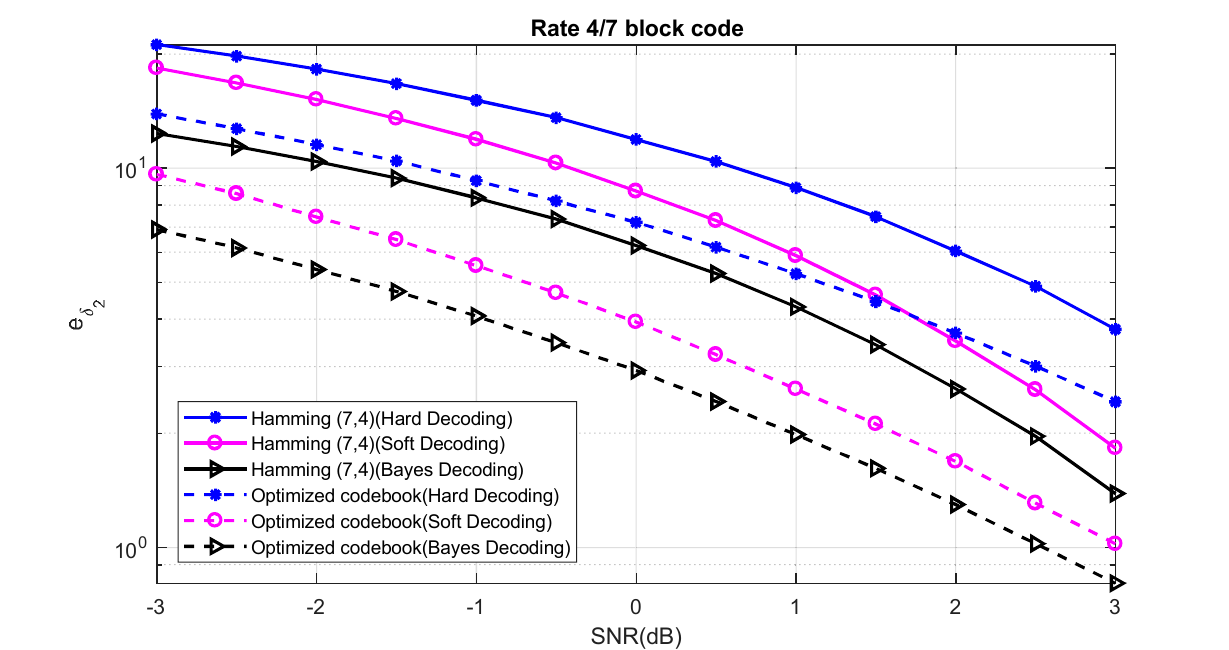}}
\caption{$e_{\delta_2}$ versus SNR for the rate 4/7 optimized code compared with Hamming (7,4) code.}\label{fig:hamming-l2}
\end{figure}

To test the performance of the optimized codebooks and the various decoding methods, we simulated the system using $10^6$ random symbols encoded with the codebook, modulated with baseband BPSK and transmitted through an AWGN channel at various signal-to-noise ratios (SNR). We estimate the variance of the noise as described above by using $10^4$ samples at the receiver.
The results of $e_{\delta_1}$ and  $e_{\delta_2}$  versus SNR are shown in Figs. \ref{fig:hamming-l1} and \ref{fig:hamming-l2} using the respective optimized codebook and Bayes decoder optimized for SNR = 0 db. We observe the following in both figures:
hard decoding is worse than soft decoding which is worse than Bayes decoding. In all three decoding schemes, the optimized codebook performs better than the Hamming code. 
For $e_{\delta_1}$, soft decoding performs almost as well as Bayes decoding.
For hard decoding, the optimized codebook can be worse than the Hamming code for large SNR. This is because the codebook is tuned for a specific noise SNR, and  the performance improvement over the Hamming code decreases (and can become inferior) as the SNR deviates from the tuned SNR (of $0$ db in our examples). 
For $e_{\delta_2}$, using both an optimized codebook with Bayes decoder results in an error that is about a third smaller than Hamming (7,4) code with hard decoding.
Finally, the benefits of using the optimized codebook and Bayes decoding are more significant for $e_{\delta_2}$ than for $e_{\delta_1}$.

\subsection{Block code of rate 3/8} \label{sec:example1}
\begin{figure}[htbp]
\centerline{\includegraphics[width=5in]{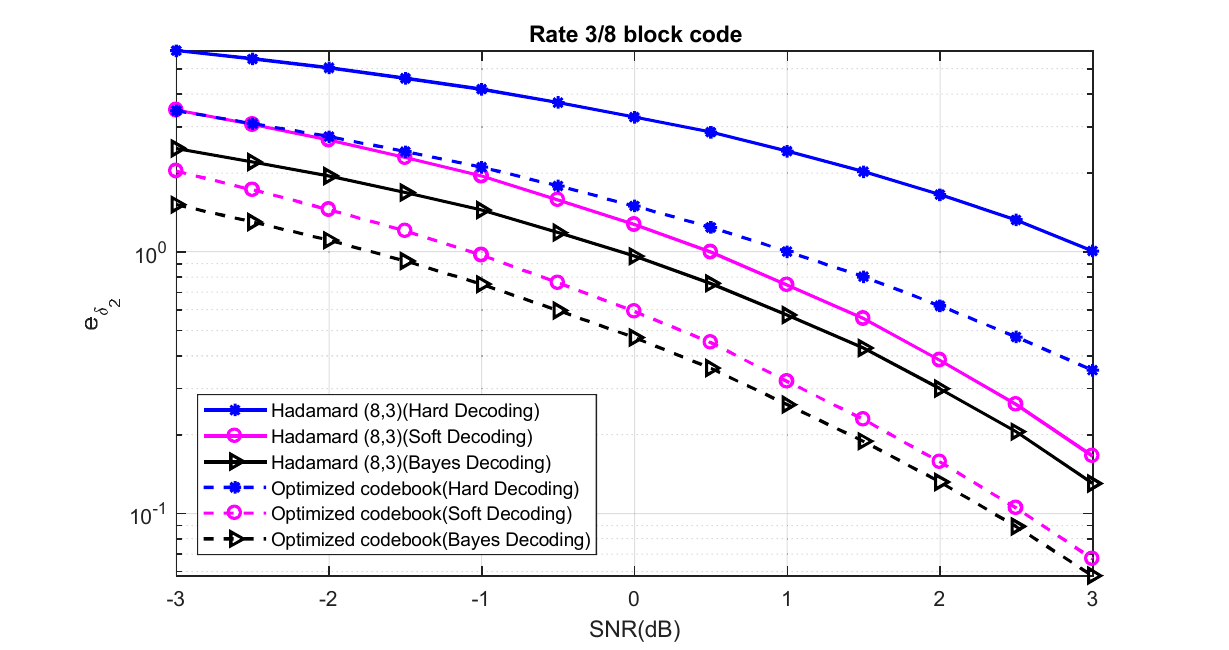}}
\caption{$e_{\delta_2}$ versus SNR for the rate 3/8 optimized code compared with Hadamard (8,3) code.}\label{fig:hadamard-l2}
\end{figure}

We repeated the same experiment with $k=3$ and $8$-bit codewords and compared the optimized codebook with the Hadamard (8,3) code. The simulation results for $\delta_2$ are shown in Fig. \ref{fig:hadamard-l2} which have similar trends as Fig. \ref{fig:hamming-l2}.

\subsection{Block code of rate 8/12} \label{sec:example1}
\begin{figure}[htbp]
\centerline{\includegraphics[width=5in]{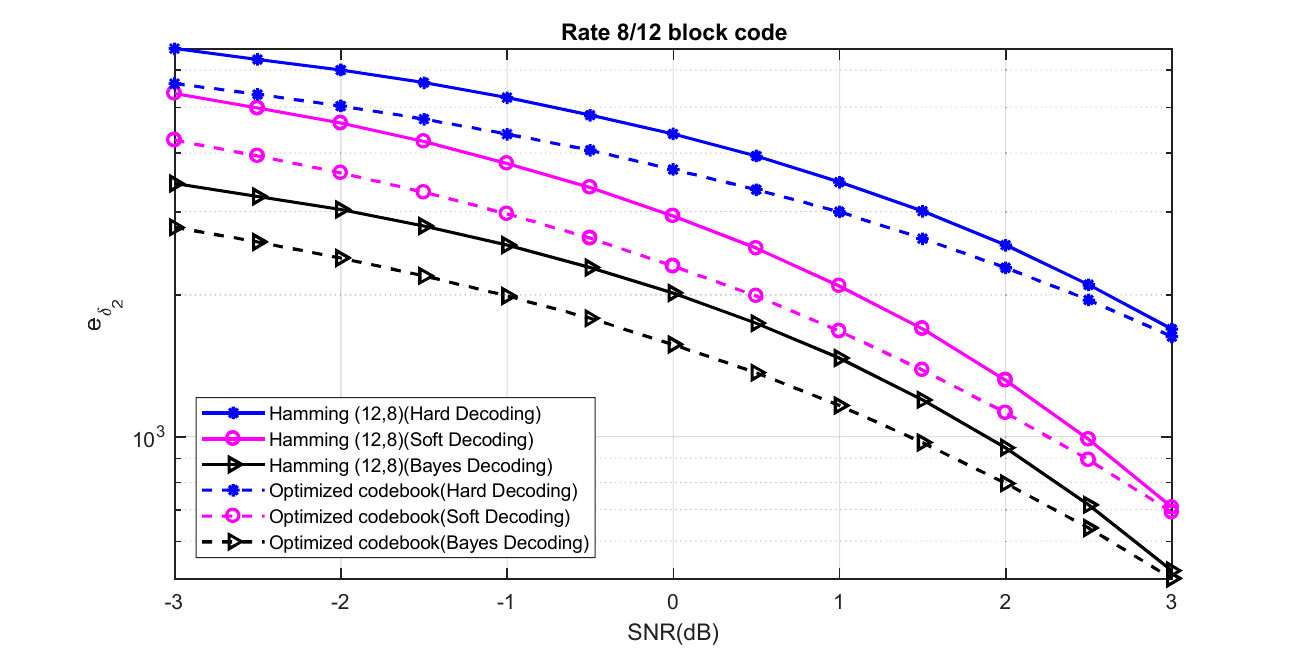}}
\caption{$e_{\delta_2}$ versus SNR for the rate 8/12 optimized code compared with Hamming (12, 8) code.}\label{fig:hamming-8-12-l2}
\end{figure}

Next, we consider a rate 8/12 code that maps 8-bit integers to 12-bit codewords and compare an optimized codebook with the modified Hamming (12, 8) code.  The simulation results for $\delta_2$ are shown in Fig. \ref{fig:hamming-8-12-l2} which again have similar trends as Fig. \ref{fig:hamming-l2}.On the other hand, because of the slow convergence time, the resulting optimized obtained after a finite stopping time is far from optimal (as we will see in Section \ref{sec:linear})

\begin{figure}[htbp]
\centerline{\includegraphics[width=5in]{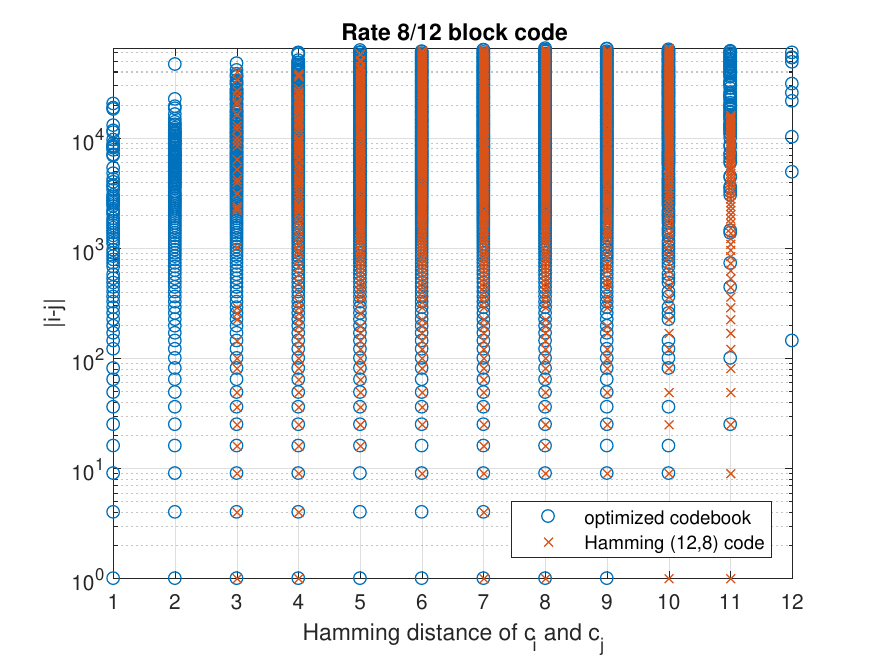}}
\caption{Hamming distance of codewords of $i$ and $j$ versus $|i-j|$ for rate 8/12.}\label{fig:hamming-12-8-codebook}
\end{figure}

Analogous to Fig. \ref{fig:l1-codebook}, we plot the Hamming distance of codewords of $i$ and $j$ versus $|i-j|$ in Fig. \ref{fig:hamming-12-8-codebook} and again the Hamming code is clustered around a few Hamming distance.

\subsection{Linear block codes} \label{sec:linear}
\begin{figure}[htbp]
\centerline{\includegraphics[width=5in]{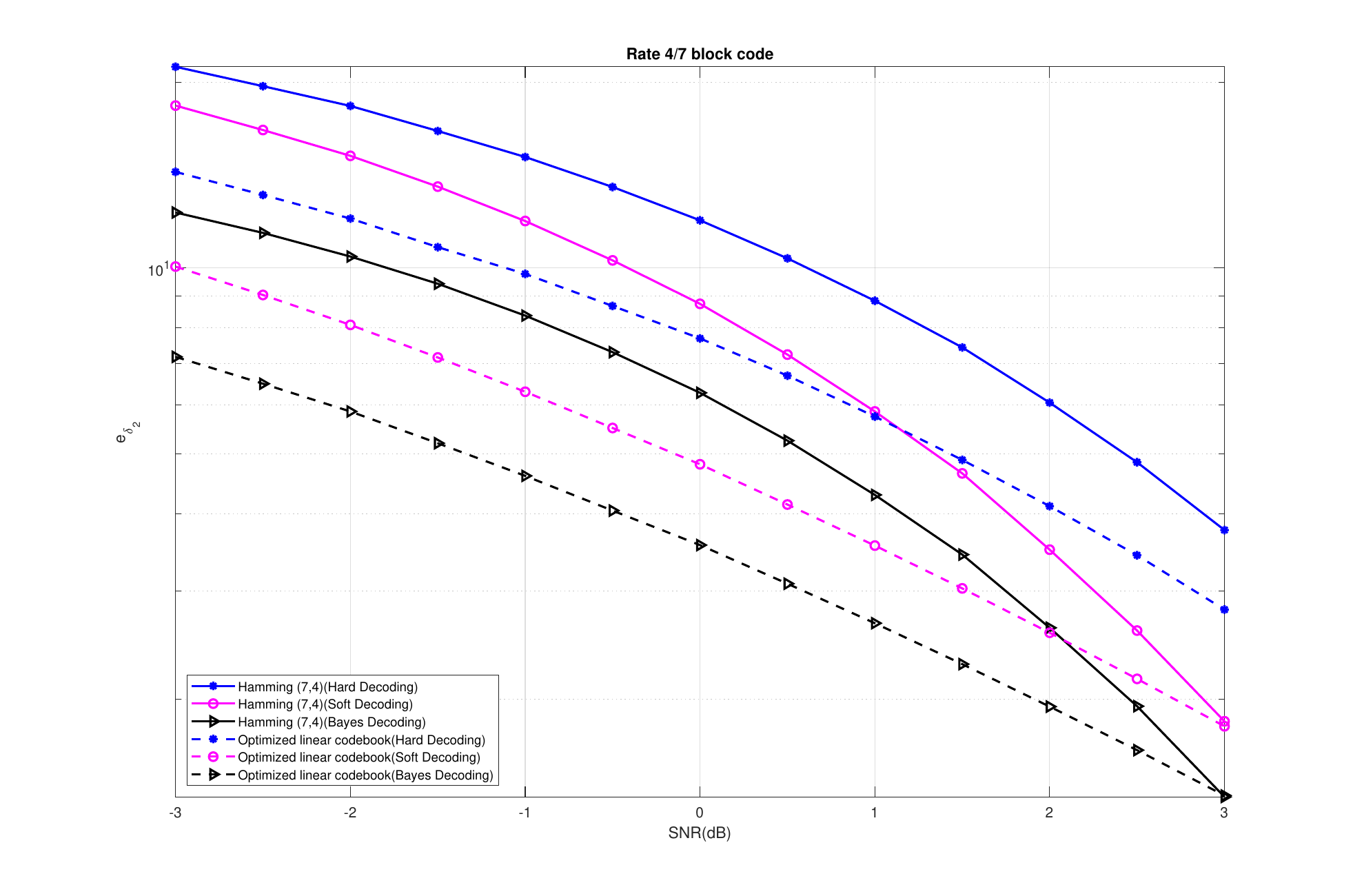}}
\caption{$e_{\delta_2}$ versus SNR for the rate 4/7 optimized linear code compared with Hamming (7,4) code.}\label{fig:hamming-l2-linear}
\end{figure}

Searching the entire set of $(n,k)$ block codes for an optimal codebook corresponds to searching the space of $2^k$ by $n$ $0$-$1$ matrices. In this section we restrict ourselves to the set of {\em linear} block codes \footnote{of which the Hamming codes are an example of.} which are defined by a $k$ by $n$ {\em generator} $0$-$1$ matrix $G$. In this case, for a symbol expressed as a $1$ by $k$ row vector $x$ of $0$'s and $1$'s, the corresponding codeword is the $1$ by $n$ row vector given by $xG$ where the arithmetic is done in the field $\mathbb  F_2$. Since the space of generator matrices is much smaller than the space of block codes, the search can be done more efficiently and the suboptimality of searching the subspace of linear codes is compensated by faster convergence, in particular when the size of the codebook is large. 
We find an optimized linear block code by using a genetic algorithm to search the space of generator matrices. We show in Fig. \ref{fig:hamming-l2-linear} the rate $4/7$ linear block code optimized for $\delta_2$. Compared with Fig. \ref{fig:hamming-l2}, we see that this code performs worse that the code optimized over all codebooks, but it still performs better than the Hamming (7,4) code over a range of SNR.
The generator matrix $G$ (in nonstandard form) is given by:
$$
G = \left(\begin{array}{ccccccc} 0 & 1 & 0 & 1 & 0 & 1 & 1\\ 0 & 0 & 0 & 0 & 1 & 0 & 0\\ 0 & 0 & 1 & 0 & 0 & 0 & 0\\ 1 & 0 & 0 & 0 & 0 & 0 & 0 \end{array}\right)
$$

On the other hand, when we consider the optimized linear block code for rate $8/12$ and comparing it with a Hamming (12,8) code, the conclusions are different. Because of the size of the codebook, searching the codebook space is much slower and after a similar number of generations of running the genetic algorithm, optimizing the linear code (Fig. \ref{fig:hamming-l2-linear}) results in a better codebook than searching over the codebook space (Fig.  \ref{fig:hamming-8-12-l2}). This indicates that the codebook obtained by searching the codebook space is far from optimal.The generator matrix of the optimized linear code is given by:

$$
G = \left(\begin{array}{cccccccccccc} 0 & 1 & 1 & 0 & 1 & 0 & 0 & 1 & 0 & 1 & 0 & 0\\ 1 & 0 & 0 & 1 & 0 & 1 & 1 & 1 & 1 & 0 & 1 & 1\\ 1 & 0 & 0 & 0 & 0 & 0 & 1 & 0 & 1 & 0 & 1 & 1\\ 1 & 0 & 0 & 0 & 0 & 0 & 0 & 0 & 1 & 0 & 1 & 0\\ 1 & 0 & 0 & 1 & 0 & 0 & 1 & 0 & 1 & 0 & 0 & 1\\ 1 & 0 & 0 & 1 & 0 & 0 & 1 & 0 & 1 & 0 & 1 & 1\\ 0 & 0 & 0 & 0 & 0 & 0 & 0 & 0 & 0 & 0 & 0 & 1\\ 1 & 0 & 0 & 0 & 0 & 0 & 0 & 0 & 0 & 0 & 1 & 0 \end{array}\right)
$$

\begin{figure}[htbp]
\centerline{\includegraphics[width=5in]{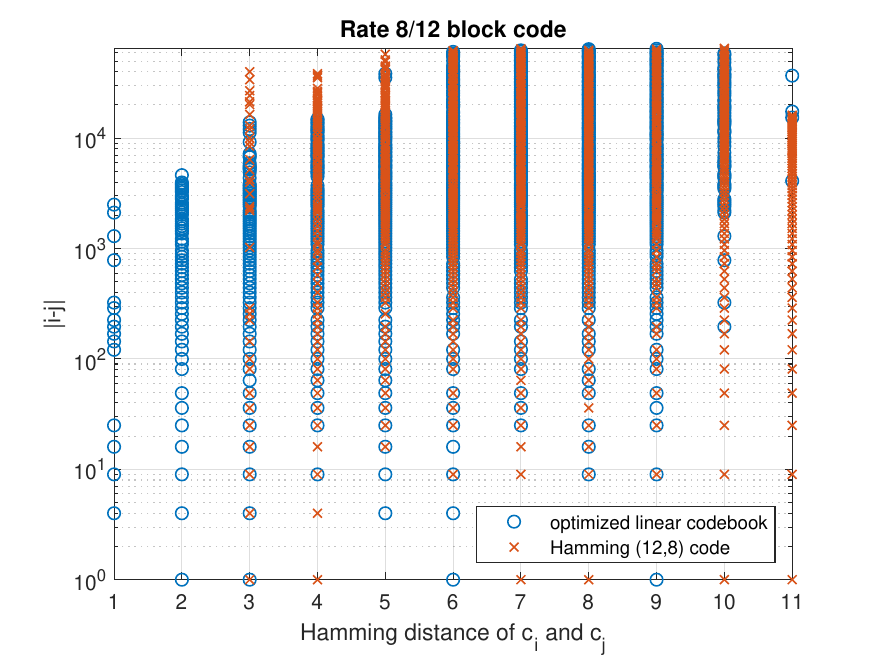}}
\caption{Hamming distance of codewords of $i$ and $j$ versus $|i-j|$ for rate 8/12.}\label{fig:hamming-12-8-linear-codebook}
\end{figure}

In Fig. \ref{fig:hamming-12-8-linear-codebook} we also plot the Hamming distance of two codewords $c_i$ and $c_j$ versus $|i-j|$ for the optimized linear rate $8/12$ code. When compared with Fig. \ref{fig:hamming-12-8-codebook} this is another indication that the optimized linear code is better than the optimized codebook due to the faster convergence and the finite stopping time.

\begin{figure}[htbp]
\centerline{\includegraphics[width=5in]{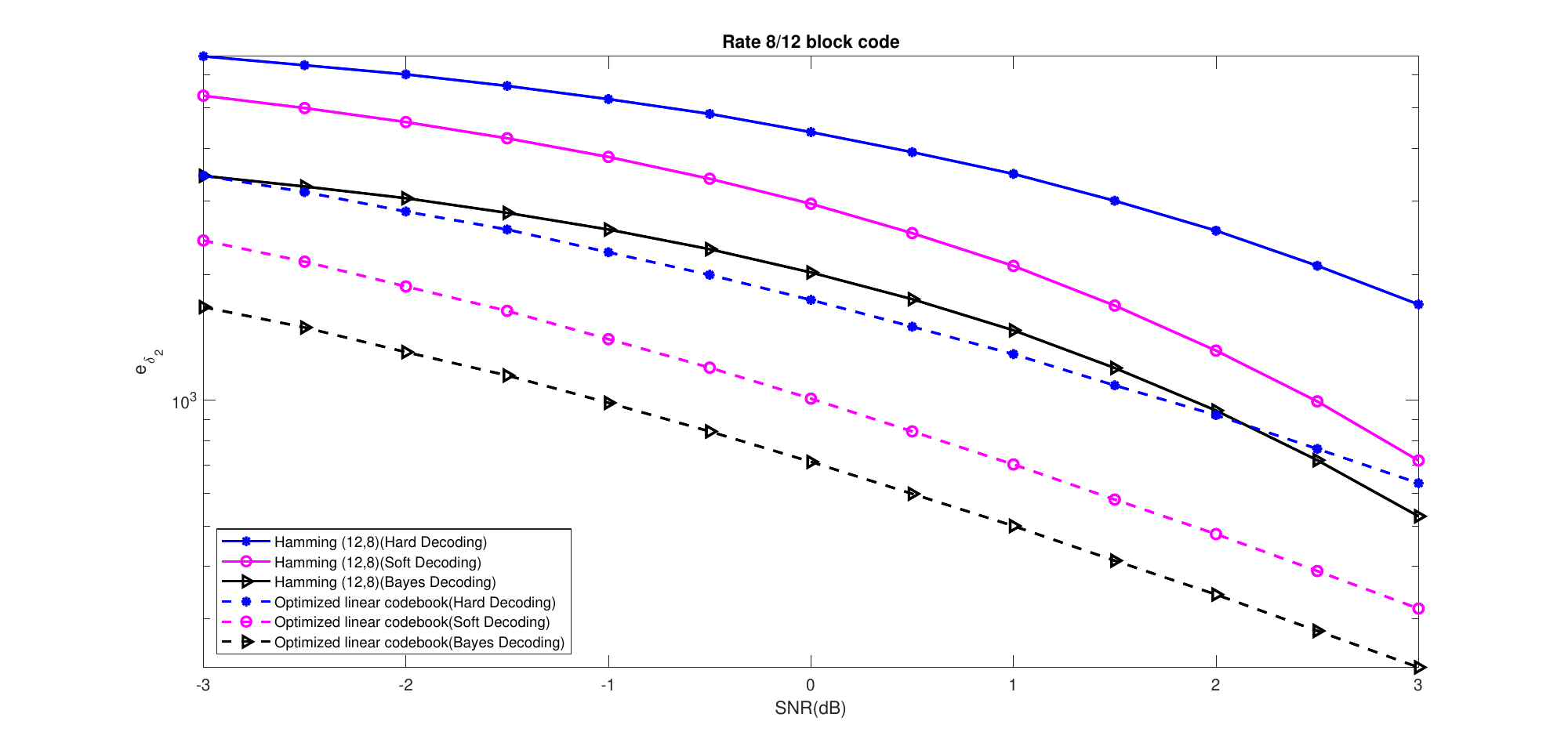}}
\caption{$e_{\delta_2}$ versus SNR for the rate 8/12 optimized linear code compared with Hamming (12,8) code.}\label{fig:hamming-12-8-l2-linear}
\end{figure}

\section{Signed integers in two's complement format}
\begin{figure}[htbp]
\centerline{\includegraphics[width=5in]{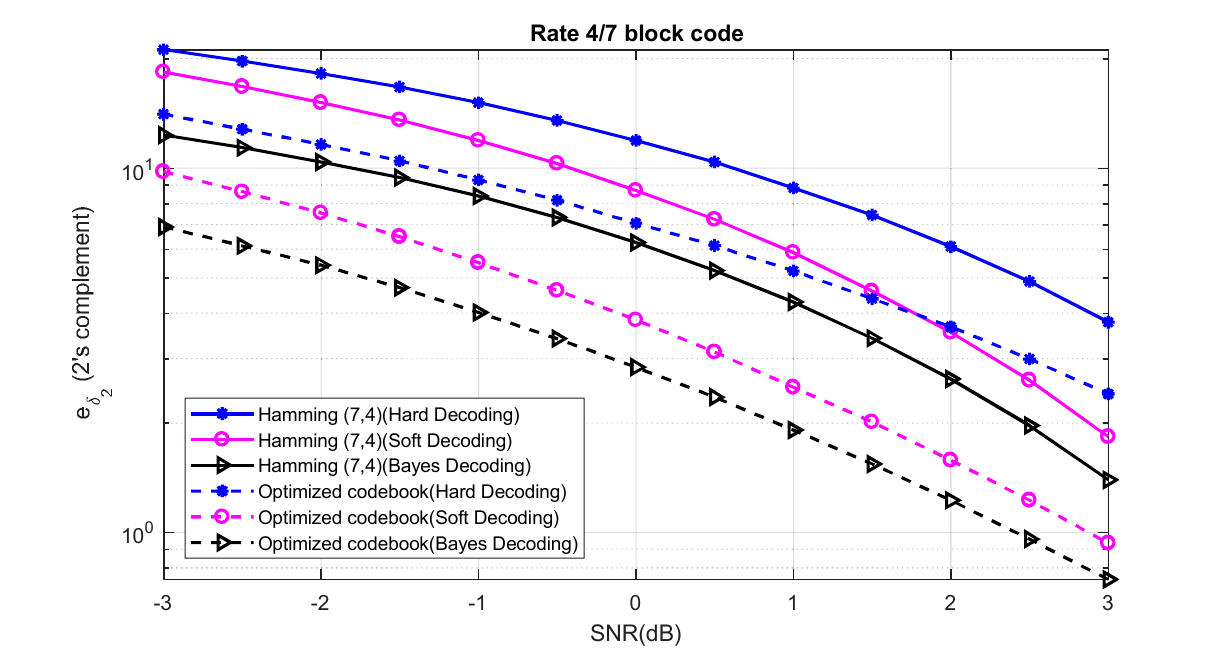}}
\caption{$e_{\delta_2}$ versus SNR where the source symbols are represented in 2's complement format.}\label{fig:hamming-l2-complement}
\end{figure}

So far we have considered nonnegative integers represented as bitstrings. The same approach can be applied to signed integers. Of course, the optimal codebook can be very different in this case. Consider the case of representing positive and negative integers $-2^{k-1} \leq i <  2^{k-1} $ in $k$-bit  2's complement format. Note that in this case the Hamming distance between $0$ and $-1$ is $k$.  For the case $k=4$, rate $4/7$, the optimized codebook for $e_{\delta_2}$ is: 

$$\left(\begin{array}{ccccccc} 0 & 0 & 0 & 1 & 0 & 1 & 1\\0 & 0 & 0 & 0 & 1 & 1 & 1\\0 & 0 & 0 & 1 & 1 & 1 & 1\\1 & 0 & 0 & 1 & 1 & 1 & 1\\0 & 1 & 0 & 1 & 1 & 1 & 1\\1 & 1 & 0 & 0 & 1 & 1 & 1\\1 & 1 & 0 & 1 & 1 & 1 & 1\\1 & 1 & 1 & 1 & 1 & 1 & 1\\1 & 1 & 1 & 0 & 0 & 0 & 0\\0 & 1 & 1 & 0 & 0 & 0 & 0\\1 & 0 & 1 & 0 & 0 & 0 & 0\\0 & 0 & 1 & 0 & 0 & 0 & 0\\0 & 0 & 1 & 1 & 0 & 0 & 0\\0 & 0 & 0 & 0 & 0 & 0 & 0\\0 & 0 & 0 & 1 & 0 & 0 & 0\\0 & 0 & 0 & 1 & 1 & 0 & 0\end{array}\right)$$
where the first, $8$-th, $9$-th, and last row corresponds to $s = 0$, $s = 7$, $s=-8$, and $s = -1$, respectively.
The numerical results are shown in Fig. \ref{fig:hamming-l2-complement} which again are very similar trendwise to Fig. \ref{fig:hamming-l2}.

\section{Comparison with related work}
In unequal error protection (UEP), different classes of messages are assigned different priorities and different ECC methods are applied to each class \cite{Borade2009,trott:uep:1996}. While there are similarities in the sense that data of higher importance should have more protection against communication errors, there are some differences as well.
In most prior UEP frameworks, the data is classified into multiple priority classes and different ECC schemes are applied independently to each class and the errors between the different classes do not trade off against each other.
In the current approach the bits have different significance and they all contribute to the same objective (e.g. contributing to the value of an integer) and thus their error correction should be treated holistically. To do this, we formulate an optimization problem where the objective function combines the various bits with different significances and thus such a trade off is possible. 
Furthermore, a Bayes optimal decoder is proposed here which for certain objective functions has a simple form and implementation.
While ML approaches have been applied to classical ECC code design (e.g.  \cite{haas:ecc:2007}), it has not been applied to code design with different bit significance. Also, in contrast to  \cite{haas:ecc:2007}, the current approach can be data-driven and use information extracted from empirical channel data (see Section \ref{sec:results}).

\section{A new number format for computer memory}
Soft error in semiconductors can cause the data in memory devices to randomly undergo bit flipping errors \cite{Slayman:2011}. This problem will be of more concern in emerging technology such as memristive memories \cite{Pouyan:2014}, quantum memories \cite{gouet:2012}, chemical and biological memories \cite{Ceze2019} where the probability of a soft error can be unpredictable and large. Furthermore, when data is stored in memory for an extended amount of time, the error rate will also increase.
Error correction circuitry can be added to memory chips, but this adds additional cost, energy and latency. In most numerical computing applications, the tolerance to error is low, and thus memory components have very low error rate. In this section, 
we consider applications such as deep learning where such errors are tolerated as the input is noisy and imprecise. In addition, significant noise is added to stochastic rounding schemes to improve the performance of low precision deep learning \cite{Gupta2015} and ODE solvers \cite{Hopkins2020}. Similarly, in stochastic computing \cite{Alaghi2013,liu:2020}, the data is encoding in stochastic pulses. In these applications, memory elements with very high error rate can be tolerated and the goal of this section is to find encoding schemes for such memory elements where the mean error is minimized. 

Define $N_k$ as the set of integers $\{0,1,\cdots, 2^k-1\}$.
We consider the problem of storing numbers in $N_k$ into $k$-bits of storage. Let us denote the set of $k$-bit binary strings as $S_k$. The storage encoding corresponds to a bijection $\sigma$ from $N_k$ to $S_k$, i.e. there are $k!$ different encodings $\sigma$ possible.
The canonical encoding, which we denote as $\sigma_*$, is defined as mapping each member of $N_k$ to its representation in binary, i.e. $\sigma_*(0) = 00\cdots 00$, $\sigma_*(1) = 00\cdots 01$, etc. and is equal to $b_k$ defined in Section \ref{sec:setup}.
With this canonical bijection, we can use the notation $N_k$ and $S_k$ interchangeably, depending on context.

Let us assume that each bit in a $k$-bit memory unit can flip independently according to a probability $p \in [0,1]$. The value of $p$ is unknown, but follows a distribution $X_p$.
Consider a number $x\in N_k$ drawn from a probability distribution $X$ with support in $N_k$ and stored in the memory unit as a bitstring $\sigma(x)$.
Due to the error in the memory unit, the bitstring $\sigma(x)$ will change to a bitstring $s\in S_k$ where each bit changes parity with probability $p$ and its corresponding number $\sigma^{-1}(s)$ will be a random variable $Y$ (that depends on $p$). Since $p$ is unknown, we will not apply any processing to $\sigma^{-1}(s)$ to recover $x$. We assume that $X$ and $X_p$ are independent.
Define $E_{x,p} = E_Y(d(x,Y)|X_p=p)$, $E_p =  E_{X,Y}(d(x,Y)|X=x,X_p=p)$.
The mean error is given by $E= E_{X,Y,X_p}(d(x,Y)|X=x,X_p=p) = \int_{0}^1 \int_{0}^1 p_{X_p}(q)p_X(x) E_{x,p} dxdq$.
In the sequel we assume both $X$ and $X_p$ to be the uniform distribution on $[0,1]$. First let us consider the case $d(x,y) = \|x-y\|^2$.

\begin{figure}[htbp]
\centerline{\includegraphics[width=\textwidth]{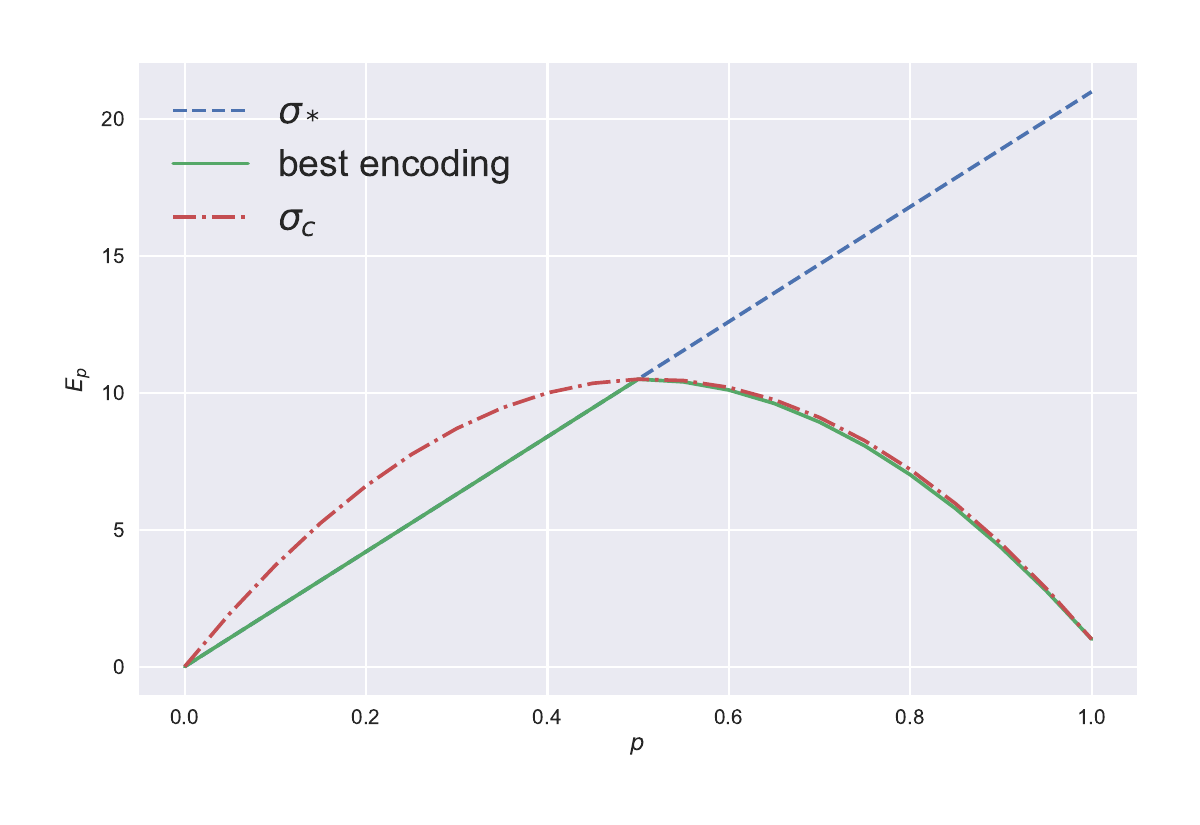}}
\caption{Canonical vs best encoding vs $\sigma_c$ ($k=3$) for $d(x,y) = \|x-y\|^2$.}\label{fig:encode-k3}
\end{figure}

Numerical experiments indicate that when $p\leq 0.5$, the canonical encoding $\sigma_*$ is minimal among all encodings. For instance, for the case $k=3$, we show in Fig. \ref{fig:encode-k3}
the error $E_p$ for the standard encoding, and the error $E_p$ for the optimal encoding (among all $8!=40320$ permutations) at each $p$.
It shows that for $p> 0.5$, the optimal encoding can have substantially lower $E_p$ than the canonical encoding $\sigma_*$. Since $p$ is unknown, we want to find a single encoding $\sigma$ such that
$E$ (i.e. $E_{X_p}(E_p)$) is minimal.

This figure suggests that if $X_p$ has support in $[0,\frac{1}{2}]$, then the canonical encoding has minimal mean error. However, if $p$ can range over all of $[0,1]$, then there are other encodings for which $E$ is lower than the canonical encoding. 
Based on these numerical experiments,  we conjecture the following:

\begin{conjecture} 
If $p\leq 0.5$, the canonical encoding has minimal error among all encodings. There exists an encoding with minimal error among all encodings and all $p\geq 0.5$.\label{conj:encode}
\end{conjecture}

The minimal encoding for $p\geq 0.5$ as conjectured to exist by Conjecture \ref{conj:encode} is difficult to compute as it needs to be optimal for all $p\geq 0.5$. Even if we restrict to a small set of $p$ for large $k$ finding the minimal encoding is hard to find due to combinatorial explosion. One of the purposes of this section is to introduce an encoding that is easy to define and implement that appears to be near optimal. 
In particular, consider the encoding $\sigma_c: N_k\rightarrow N_k$, where 
$\sigma_c (2n) = n$ and $\sigma_c(2n+1) = M-n$ where $M=2^k-1$. Tables \ref{tbl:sumdiff} and \ref{tbl:prod} show the formulas on sums and products of numbers encoded in this format.

\begin{table}[htb]
\begin{center}
\begin{tabular}{|c|c|c|}
\hline
 & $\sigma_c(a+b)$ & $\sigma_c(a-b)$  \\ \hline\hline
 $a$, $b$ even &  $\sigma_c(a)+\sigma_c(b)$ &  $\sigma_c(a)-\sigma_c(b)$  \\  \hline
 $a$ even, $b$ odd & $\sigma_c(b)-\sigma_c(a)$ & $\sigma_c(a+1)-\sigma_c(b+1)$     \\ \hline
  $a$ odd, $b$ even & $\sigma_c(a)-\sigma_c(b)$ & $\sigma_c(a)+\sigma_c(b)$   \\ \hline
   $a$, $b$ odd & $2M+1-\sigma_c(a)-\sigma_c(b)$ & $\sigma_c{b}-\sigma_c(a)$      \\ \hline
\end{tabular}
\end{center}
\caption{Sum and difference formulas for $\sigma_c$.}\label{tbl:sumdiff}
\end{table}

\begin{table}[htb]
\begin{center}
\begin{tabular}{|c|c|}
\hline
 & $\sigma_c(ab)$ \\ \hline\hline
 $a$, $b$ even & $2\sigma_c(a)\sigma_c(b)$ \\  \hline
 $a$ even, $b$ odd  & $\sigma_c(a)(2M-2\sigma_c(b)+1)$     \\ \hline
 $a$, $b$ odd & $(2M+1)(\sigma_c(a)+\sigma_c(b)-M)-2\sigma_c(a)\sigma_c(b)$\\ \hline

\end{tabular}
\end{center}
\caption{Product formulas for the encoding $\sigma_c$.}\label{tbl:prod}
\end{table}

These formulas show that addition, subtraction and multiplication can be performed on fixed point nonnegative numbers encoded in this format  via relatively straightforward modification of standard arithmetic circuits. This means that an arithmetic logic unit (ALU) for this number format should have similar hardware complexity as that for the traditional number format.

\begin{figure}[htbp]
\centerline{\includegraphics[width=\textwidth]{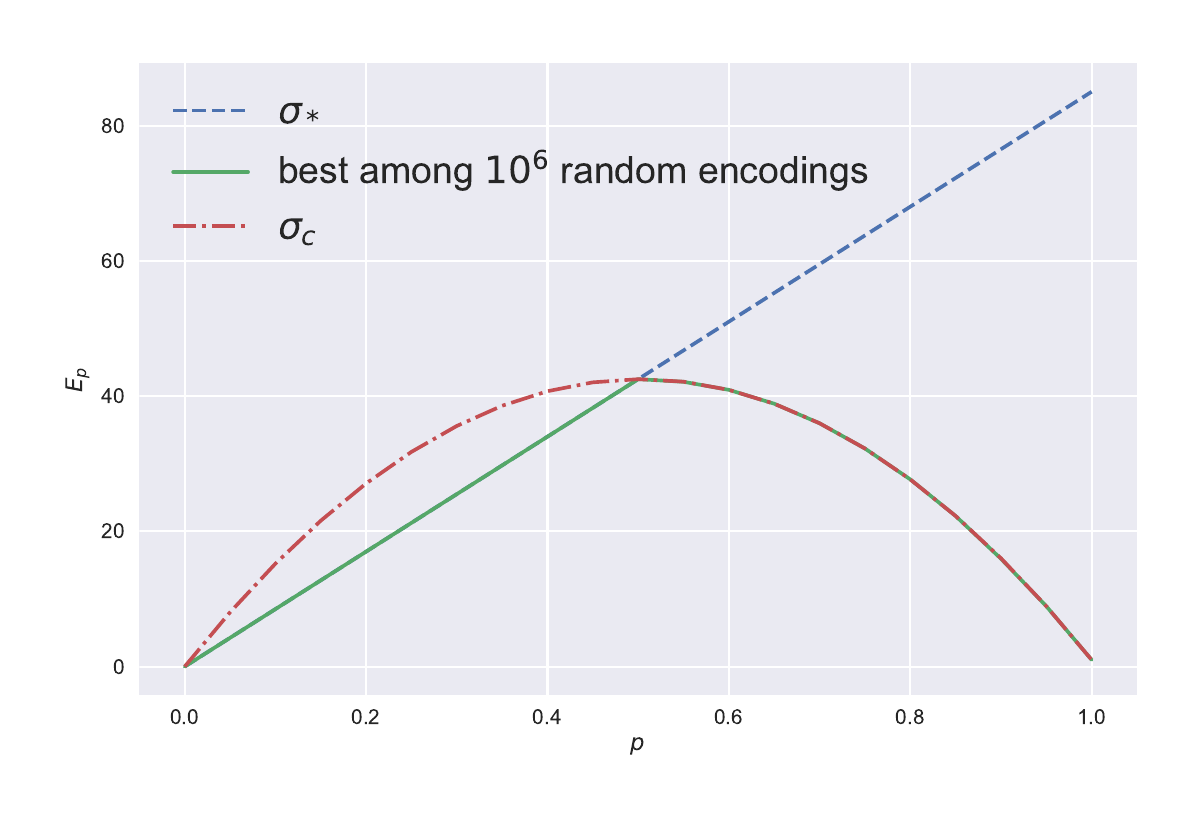}}
\caption{Canonical vs best among $10^6$ encodings vs $\sigma_c$ ($k=4$) for $d(x,y) = \|x-y\|^2$.}\label{fig:encode-k4}
\end{figure}

Fig. \ref{fig:encode-k3} shows that the error $E_p$ for $\sigma_c$ is close to optimal for $k=3$ and $p \geq 0.5$.
Fig. \ref{fig:encode-k4} shows for $k=4$ and each $p$ the minimal $E_p$ among $10^6$ random bijective encodings. In this case this minimal $E_p$ is indistinguishable from the error for $\sigma_c$ for $p\geq 0.5$.

\begin{figure}[htbp]
\centering
\centerline{\includegraphics[width=\textwidth]{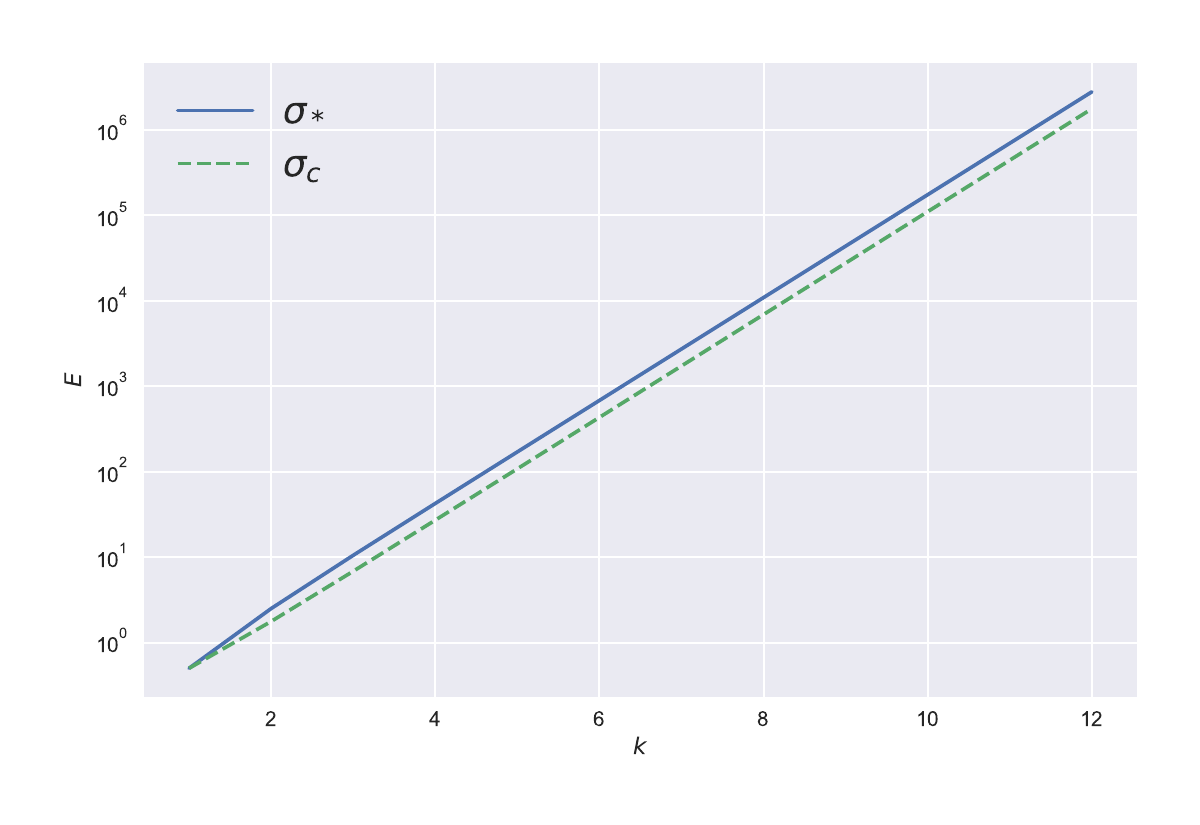}}
\caption{Mean error $E$ for the canonical encoding vs  $\sigma_c$ for $d(x,y) = \|x-y\|^2$.}\label{fig:efigure}
\end{figure}

Figure \ref{fig:efigure} shows that the mean error $E$ is less for $\sigma_c$ than $\sigma_*$ for various values $k$ and appears asymptotically to differ by a multiplicative constant $\approx 1.579$. This indicates that using $\sigma_c$ rather than the canonical encoding $\sigma_*$ will result in a lower average error when the bit-flipping probability $p$ is unknown and can range from $0$ to $1$.

\begin{figure}[htbp]
\centerline{\includegraphics[width=\textwidth]{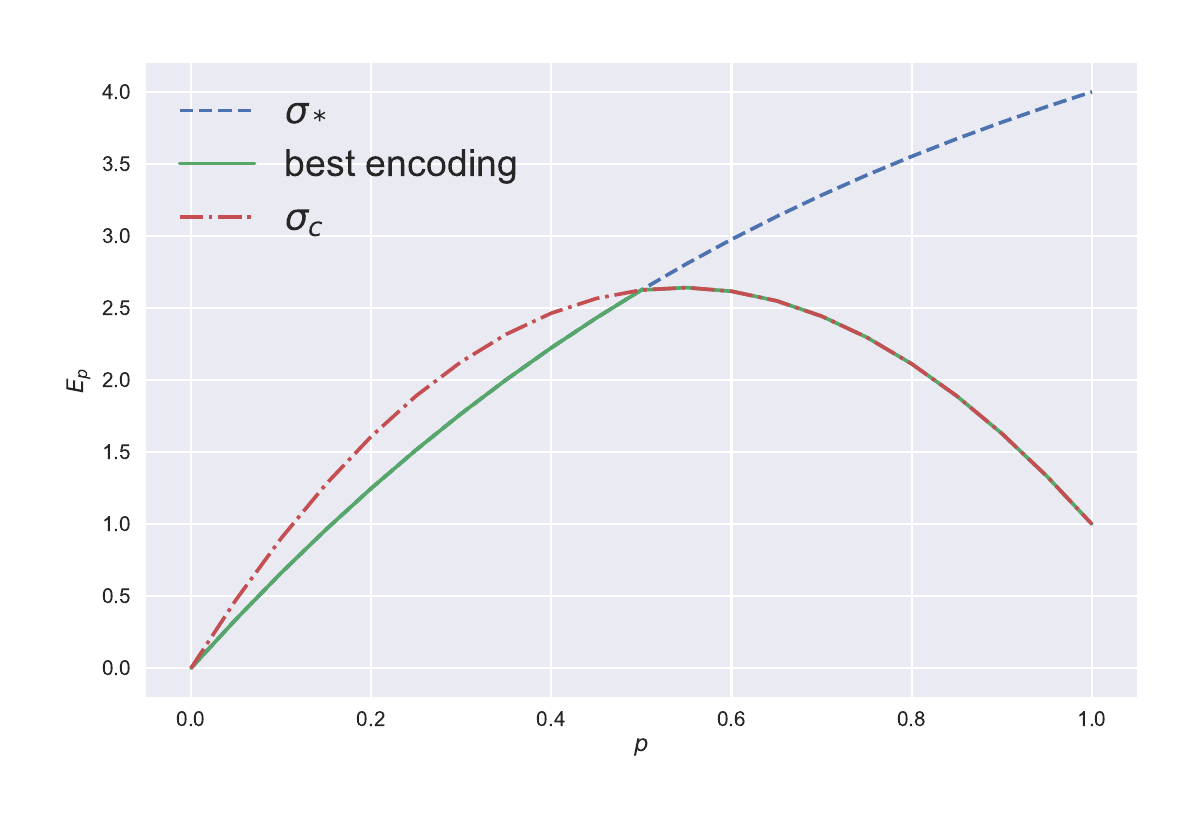}}
\caption{Canonical vs best encoding vs $\sigma_c$ ($k=3$) for $d(x,y) = \|x-y\|$.}\label{fig:encode-k3l1}
\end{figure}

\begin{figure}[htbp]
\centering
\centerline{\includegraphics[width=\textwidth]{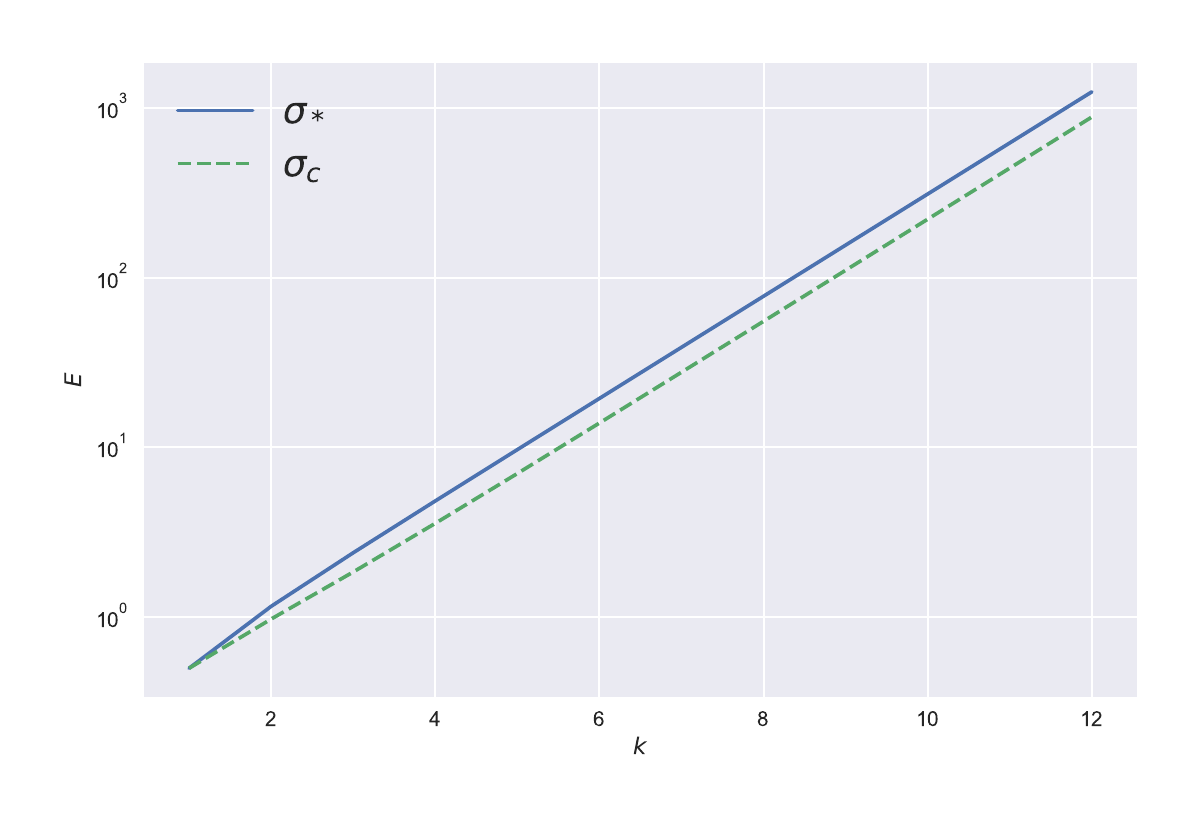}}
\caption{Mean error $E$ for the canonical encoding vs  $\sigma_c$ for $d(x,y) = \|x-y\|$.}\label{fig:efigurel1}
\end{figure}

The corresponding figures for $d(x,y) = \|x-y\|$ is shown in Figs. \ref{fig:encode-k3l1} and \ref{fig:efigurel1}.
Numerical results show similar behavior when $N_k$ are the set of integers ${-2^{k-1},\cdots, 0, 1,\cdots, 2^{k-1}-1}$ in $k$-bit 2's complement format.

\section{Concluding remarks}
We proposed a framework for error correction coding that takes into account the difference in bit significance in the source symbols by using an appropriate error metric and minimizing it using a Bayes decoder and an optimized codebook derived from iterative improvement search techniques. We show that the Bayes decoder performs better than standard soft and hard minimum distance decoding and that the optimized codebook performs better than classical linear block codes such as Hamming codes. 

The error metric based on the difference $|i-j|$ is similar to assigning an exponential weight $2^d$ to the $d$-th bit. The same approach can be applied to  other ways of assigning significance to the various bits in the source bit stream by defining $\delta$ appropriately. In addition, even though we have been discussing the bits to be independent in terms of their significance, the loss function $L$ described above is more general in the sense that it compares two different source symbols as a whole, not just comparing them bitwise, and thus can take into account the correlation among bits. 

Furthermore, we have considered the AWGN channel model in this paper, but the Bayes estimator can be defined and the codebook can be optimized for other channel models (such as BSC) as well. In such cases, deep neural networks can be used to estimate the unknown channel and noise characteristics \cite{Ye2018}. For instance, a data-driven approach can be used to estimate the probabilities  $p_c(\Phi(s_i)|\Phi(s_j))$  in the definition of $v$. 

In the examples above, the symbol space $S$ has size $2^k$. This is not a necessary requirement; the optimized codebook can be built to map any number of $m$ symbols to $n$-bit codewords, resulting in a rate $\log_2(m)/n$ code, thus providing a more flexible tradeoff between rate and distortion. 

The cases of a nonuniform prior distribution for the source symbols or the addition of source (entropy) coding result in a more complicated source symbol posterior distribution, Bayes estimator and the function $v$ used in the heuristic. In addition, the function $v$ can be amended to optimize for multiple noise SNR's. Furthermore, noise models and parameters can be estimated by sending intermittent probe bits through the channel \cite{ahn:channel-probe:2002}. Finally, different codebooks can be used (and the codebook choice communicated between the transmitter and the receiver) based on the (estimated) noise models and/or parameters. 

In order to use a single codebook minimizing a specific objective function $v$, we assume the same semantics for the data that is being transmitted. This is a reasonable assumption especially for the data transfer of large scale data sets where nearly all of the data are of the same type. 
For instance, in the transmission of speech or other multimedia information, various header information is sent to allow the receiver to know what type of data (e.g. DCT coefficients, ASCII, etc.) are being sent and different codebooks can be used depending on the type of data that is being sent.
These topics will be discussed in a future paper.

Finally, for memory storage where the bit flipping probability can be large and unknown, a novel number encoding format is proposed that has a smaller expected error than the traditional encoding format.

\end{document}